\documentclass[aps,prl,reprint,groupedaddress,amsmath,amssymb]{revtex4-1}
\usepackage{graphicx}
\begin{document}

\title{Negative refractive index induced by percolation in disordered metamaterials}

\author{Brian A. Slovick}\email{Corresponding author: brian.slovick@sri.com}
\affiliation{Applied Optics Laboratory, SRI International, Menlo Park, California 94025, United States}
\begin{abstract}
An effective medium model is developed for disordered metamaterials containing a spatially random distribution of dielectric spheres. Similar to effective medium models for ordered metamaterials, this model predicts resonances in the effective permeability and permittivity arising from electric- and magnetic-dipole Mie resonances in the spheres. In addition, the model predicts a redshift of the electric resonance with increasing particle loading. Interestingly, when the particle loading exceeds the percolation threshold of 33\%, the model predicts that the electric resonance overlaps with the magnetic resonance, resulting in a negative refractive index.
\end{abstract}
\maketitle

Metamaterials, inhomogeneous composite materials that behave macroscopically as homogeneous materials, are of great fundamental and technological interest. The ability to tailor the macroscopic electromagnetic parameters, namely the permittivity and permeability, by changing the geometry and arrangement of the included elements has led to a number of important material designs and applications \cite{Engheta2006}. Most notable among these are materials with negative refractive index \cite{Smith2000,Shalaev2007}, which have both negative permittivity and permeability. Negative index materials exhibit unusual properties such as negative refraction and exponential growth of evanescent near fields, which have been utilized for applications such as cloaking \cite{Schurig2006} and superresolution imaging \cite{Pendry2000}.

To date, the metamaterial literature has focused primarily on designs where the included elements are arranged in a periodic lattice \cite{Smith2005}. This is largely due to the ease of computation and suppression of scattering, as disorder of the lattice leads to spatial inhomogeneities. However, fabrication of periodic metamaterials on a large scale can be cost prohibitive, particularly for metamaterials operating in the optical band, where the lattice dimensions are submicron and require the use of electron-beam lithography \cite{Boltasseva2008,Burckel2010}. Self-assembly based nanosphere lithography has been used to fabricate centimeter-scale optical metasurfaces \cite{Moitra2015}, but this approach is not scalable to much larger areas.

Disordered metamaterials provide a scalable alternative to periodic structures, provided that the scattering can be managed and their performance can be predicted. However, relatively few works have focused on disordered metamaterials. These limited studies have shown that size \cite{Zharov2005,Gorkunov2006,Gollub2007,Andryieuski2016} and positional \cite{Aydin2004,Helgert2009,Papasimakis2009,Singh2009,Muhlig2011,Albooyeh2012,Rico-Garcia2012,Savo2012,Albooyeh2014,Moitra2014} disorder of a metamaterial's resonator elements lead to resonance broadening and effective scattering loss. While these works provide important insights, they overlook one of the most important aspects of disorder: percolation. Percolation occurs in random composites at high loading when the particles form a continuous path through the composite, and it can have dramatic consequences on the effective electromagnetic parameters \cite{Kirkpatrick1973,Pecharroman2000}. For example, in metal-dielectric composites, percolation leads to a metal-insulator transition \cite{Kirkpatrick1973}, which can be exploited to obtain materials with ultrahigh dielectric constant for use as supercapacitors \cite{Pecharroman2000}. However, to date percolation studies have considered composites containing subwavelength or polydisperse particles, in which Mie scattering resonances are absent or dampened. An open question is how percolation impacts composites containing monodisperse, wavelength-sized particles that support collective Mie resonances.

In this Letter, it is shown that percolation induces a negative refractive index in disordered metamaterials containing Mie-resonant particles. To this end, the classic Bruggeman percolation theory \cite{Bruggeman1935,Kirkpatrick1973} is generalized to accommodate particle sizes comparable to the wavelength. Similar to effective medium models for ordered metamaterials \cite{Lewin1947,Wu2006,Simovski2007,Slovick2014}, this model predicts resonances in the effective permeability and permittivity arising from electric- and magnetic-dipole Mie resonances in the spheres, as well as a redshift of the electric resonance as the loading of the particles increases. Interestingly, when the particle loading exceeds the percolation threshold of 33\%, the electric resonance overlaps with the magnetic resonance, resulting in a negative refractive index. The possibility of obtaining a negative index in disordered metamaterials opens the door to more scalable designs suitable for practical large-area applications.

The model is derived by considering a composite medium containing a random dispersion of spherical particles in a matrix, which itself is represented as a space-filling, polydisperse distribution of infinitesimally small spheres. Similar to previous models \cite{Stroud1978,Niklasson1981}, the effective medium is defined such that the particles and matrix spheres, embedded in the effective medium, produce zero scattering and absorption (i.e., zero extinction). In this way, the effective medium is indistinguishable from the composite medium. Taking the extinction cross sections of the particle and matrix spheres to be $\sigma_p$ and $\sigma_{m,i}$, respectively, the condition for zero extinction is
\begin{equation}
\sigma_pN_p+\sum_i \sigma_{m,i}N_{m,i}=0,
\end{equation}
where $N_p$ and $N_{m,i}$ are the number densities of particles and matrix spheres, respectively. By dividing the matrix into spheres, Mie theory can be used to calculate the matrix extinction. Requiring the sum of the extinctions to be zero implies that the extinction of the particles or matrix spheres is negative. This can be interpreted as a multiple scattering effect, in which scattered light is returned to the unscattered beam by subsequent scattering events. In this interpretation, the effective medium is the unique medium in which all light scattered by the particles is scattered back into the specular beam by the matrix. The reduction of the extinction coefficient by multiple scattering is well documented in the optical remote sensing community \cite{Wandinger1998}. The extinction cross sections can be calculated from the forward-scattering amplitude $S_j(0)$ using the optical theorem \cite{Stroud1978,Niklasson1981,Bohren1983}
\begin{equation}
\sigma_j=\frac{4\pi}{k^2} \text{Re}[S_j(0)],
\end{equation}
where $j=(m,p)$, $k=2\pi/\lambda\sqrt{\epsilon\mu}$ is the wavenumber in the effective medium, where $\lambda$ is the free-space wavelength, and $\epsilon$ and $\mu$ are the effective permittivity and permeability, respectively. The forward scattering amplitude, in turn, is related to electric and magnetic multipole Mie scattering coefficients, $a_{n,j}$ and $b_{n,j}$ by \cite{Bohren1983,Kuester2011}
\begin{equation}
S_j(0)=\tfrac{1}{2}\sum_{n=1}^\infty (2n+1)(a_{n,j}+b_{n,j}), \quad \text{where}
\end{equation}
$$
a_{n,j}=\frac{\sqrt{\epsilon_j/\epsilon}\psi'_n(kr_j)\psi_n(k_jr_j)-\sqrt{\mu_j/\mu}\psi_n(kr_j)\psi'_n(k_jr_j)}{\sqrt{\epsilon_j/\epsilon}\xi'_n(kr_j)\psi_n(k_jr_j)-\sqrt{\mu_j/\mu}\xi_n(kr_j)\psi'_n(k_jr_j)},
$$
$$
b_{n,j}=\frac{\sqrt{\mu_j/\mu}\psi'_n(kr_j)\psi_n(k_jr_j)-\sqrt{\epsilon_j/\epsilon}\psi_n(kr_j)\psi'_n(k_jr_j)}{\sqrt{\mu_j/\mu}\xi'_n(kr_j)\psi_n(k_jr_j)-\sqrt{\epsilon_j/\epsilon}\xi_n(kr_j)\psi'_n(k_jr_j)},
$$
where $r_j$ is the radius of the particle or matrix, $\epsilon_j$ and $\mu_j$ are the permittivity and permeability, $k_j=2\pi/\lambda\sqrt{\epsilon_j \mu_j}$, $\psi_n(x)=xj_n(x)$ and $\chi_n(x)=-xy_n(x)$ are the Riccati-Bessel functions where $j_n(x)$ and $y_n(x)$ are the spherical Bessel functions, $\xi_n(x)=\psi_n(x)-i\chi_n(x)$, and the primes denote differentiation with respect to the argument.

Substituting Eq. (2) into Eq. (1) and factorizing common terms, the condition for zero extinction reduces to
\begin{equation}
S_p(0)N_p+\sum_i S_{m,i}(0)N_{m,i}=0.
\end{equation}
In order to obtain a closed-form solution to Eq. (4), several approximations are made. First, only the $n=1$ dipolar terms in the expansion of $S_j(0)$ are considered. This approximation is valid as long as the spheres are subwavelength in size, as in the case of the matrix spheres, or have large refractive index relative to the background medium, as in the case of the particles. Second, the wavelength in the effective medium is assumed to be much larger than the spheres (i.e., $kr_j<<1$). In this case, the Riccati-Bessel functions can be replaced by their small-argument approximations and the electric and magnetic dipole Mie scattering coefficients reduce to \cite{Lewin1947,Kuester2011} 
\begin{equation}
a_{1,j}\approx-\frac{2i}{3}(kr_j)^3\frac{\epsilon_jF(k_jr_j)-\epsilon}{\epsilon_jF(k_jr_j)+2\epsilon}, \quad \text{and}
\end{equation}
\begin{equation}
b_{1,j}\approx-\frac{2i}{3}(kr_j)^3\frac{\mu_jF(k_jr_j)-\mu}{\mu_jF(k_jr_j)+2\mu}, \quad \text{where}
\end{equation}
$$
F(x)=\frac{2\psi_1(x)}{x\psi_1'(x)}=\frac{2(\sin{x}-x\cos{x})}{x\cos{x}+(x^2-1)\sin{x}}.
$$
With these approximations, the condition for zero extinction in Eq. (4) reduces to
\begin{eqnarray}
&&N_pr_p^3\left[\frac{\epsilon_pF(k_pr_p)-\epsilon}{\epsilon_pF(k_pr_p)+2\epsilon}+\frac{\mu_pF(k_pr_p)-\mu}{\mu_pF(k_pr_p)+2\mu}\right] +\sum_i N_{m,i}r_{m,i}^3 \nonumber \\
\times&&\left[\frac{\epsilon_mF(k_mr_{m,i})-\epsilon}{\epsilon_mF(k_mr_{m,i})+2\epsilon}+\frac{\mu_mF(k_mr_{m,i})-\mu}{\mu_mF(k_mr_{m,i})+2\mu}\right]=0.
\end{eqnarray}
Similar to the original Bruggeman model \cite{Bruggeman1935}, Eq. (7) is symmetric with respect to interchange of the particle and matrix. However, the focus of this work is on composites containing discrete, wavelength-sized particles dispersed in a continuous matrix, which can be represented as a space-filling, polydisperse distribution of infinitesimally small spheres. Thus, the matrix spheres are assumed to be much smaller than the wavelength, in which case $F(k_mr_{m,i})=1$.

Since the terms in Eq. (7) containing $\epsilon$ are independent of $\mu$, and vice versa, the solution requires their sum to be zero. Noting that $N_pr_p^3$ is proportional to the volume fraction of particles $f$, and thus $\sum_i N_{m,i}r_{m,i}^3$ is proportional to $1-f$, the solution to Eq. (7), assuming $F(k_mr_{m,i})=1$, is
\begin{equation}
f\frac{\epsilon_pF(k_pr_p)-\epsilon}{\epsilon_pF(k_pr_p)+2\epsilon}+(1-f)\frac{\epsilon_m-\epsilon}{\epsilon_m+2\epsilon}=0,
\end{equation}
\begin{equation}
f\frac{\mu_pF(k_pr_p)-\mu}{\mu_pF(k_pr_p)+2\mu}+(1-f)\frac{\mu_m-\mu}{\mu_m+2\mu}=0.
\end{equation}
Equations (8) and (9) are quadratic equations which can be solved to obtain the following closed-form expressions for the effective permittivity and permeability:
\begin{eqnarray}
&&\epsilon=\frac{1}{4} \left\{ E \pm \left[ E^2 + 8F(k_pr_p)\epsilon_p \epsilon_m \right]^{1/2} \right\},\nonumber \\
&&E=\epsilon_m(2-3f)+F(k_pr_p)\epsilon_p(3f-1), \quad \text{and}
\end{eqnarray}
\begin{eqnarray}
&&\mu=\frac{1}{4}\left\{ M \pm \left[ M^2 + 8F(k_pr_p)\mu_p \mu_m \right]^{1/2} \right\}, \nonumber \\
&&M=\mu_m(2-3f)+F(k_pr_p)\mu_p(3f-1).
\end{eqnarray}
Several key properties of the model are now highlighted. First, to ensure thermodynamic passivity, the $\pm$ signs must be chosen such that the imaginary parts of $\epsilon$ and $\mu$ are positive \cite{Landau1984,Slovick2014}. Second, for particles much smaller than the wavelength, $F(k_pr_p)=1$ and the model reduces to the classic Bruggemen mixing model \cite{Bruggeman1935,Stroud1978,Niklasson1981}. In this limit, $\mu=1$ when the materials are nonmagnetic (i.e., $\mu_p=\mu_m=1$). The classic Bruggeman model provides an accurate description of $\epsilon$ except for a small range of volume fractions near the percolation threshold \cite{Kirkpatrick1973}. Third, the effect of wavelength-sized particles, embodied in the resonant function $F(k_pr_p)$, is to multiply the effective permittivity and permeability of the particles. Near the Mie resonances, $F(k_pr_p)$ diverges, leading to effectively large values of the particle permittivity and permeability, and thus resonances in $\epsilon$ and $\mu$. Fourth, in the limit of small volume fractions where the interparticle separation is large, interactions between particles are negligible, and the ordering of the particles should have no effect. Thus, in this limit the extended Bruggeman model should reduce to the Lewin effective medium theory for a cubic array of spheres \cite{Lewin1947,Slovick2014}. For example, Lewin's expression for $\epsilon$ can be obtained by setting $1-f\approx1$ and $\epsilon\approx \epsilon_m$ in Eq. (8),
\begin{equation}
f\frac{\epsilon_pF(k_pr_p)-\epsilon_m}{\epsilon_pF(k_pr_p)+2\epsilon_m}+\frac{\epsilon_m-\epsilon}{\epsilon+2\epsilon_m}\approx0,
\end{equation}
and solving for $\epsilon$ to obtain
\begin{equation}
\epsilon=\epsilon_m \frac{1+2f\tfrac{\epsilon_pF(k_pr_p)-\epsilon_m}{\epsilon_pF(k_pr_p)+2\epsilon_m}}{1-f\tfrac{\epsilon_pF(k_pr_p)-\epsilon_m}{\epsilon_pF(k_pr_p)+2\epsilon_m}}.
\end{equation}
The same line of reasoning can be applied to obtain a similar expression $\mu$. Thus, for small volume fractions the extended Bruggeman model is equivalent to the Lewin effective medium model. It is important to note that this equivalence does not hold close to the Mie resonances where $\epsilon>>\epsilon_m$. Lastly, when $\epsilon_p$ or $\mu_p$ is large, or the particle is resonant [i.e., $F(k_pr_p)\rightarrow \infty$], a discontinuity with loading, or percolation threshold, occurs. This can be seen, for example, by taking the limit of Eq. (10) as $F(k_pr_p)\epsilon_p\rightarrow \infty$, in which case $\epsilon=0$ for $f\leq1/3$ and $\epsilon=\tfrac{1}{2}F(k_pr_p)\epsilon_p(3f-1)$ for $f>1/3$. Since a similar argument applies to the effective permeability, beyond percolation both $\epsilon$ and $\mu$ are proportional to $F(k_pr_p)$, and thus their resonances, given by the poles of $F(k_pr_p)$, coincide to produce a negative index. Also, since $F(k_pr_p)$ always diverges for certain values of $k_pr_p$, the model predicts percolation and negative index for all values of $\epsilon_p$; however, this is an artifact of the dipole approximation which implicitly assumes large permittivity contrast ($\epsilon_p/\epsilon_m\gtrsim 10$). 

\begin{figure}
\includegraphics[width=62mm]{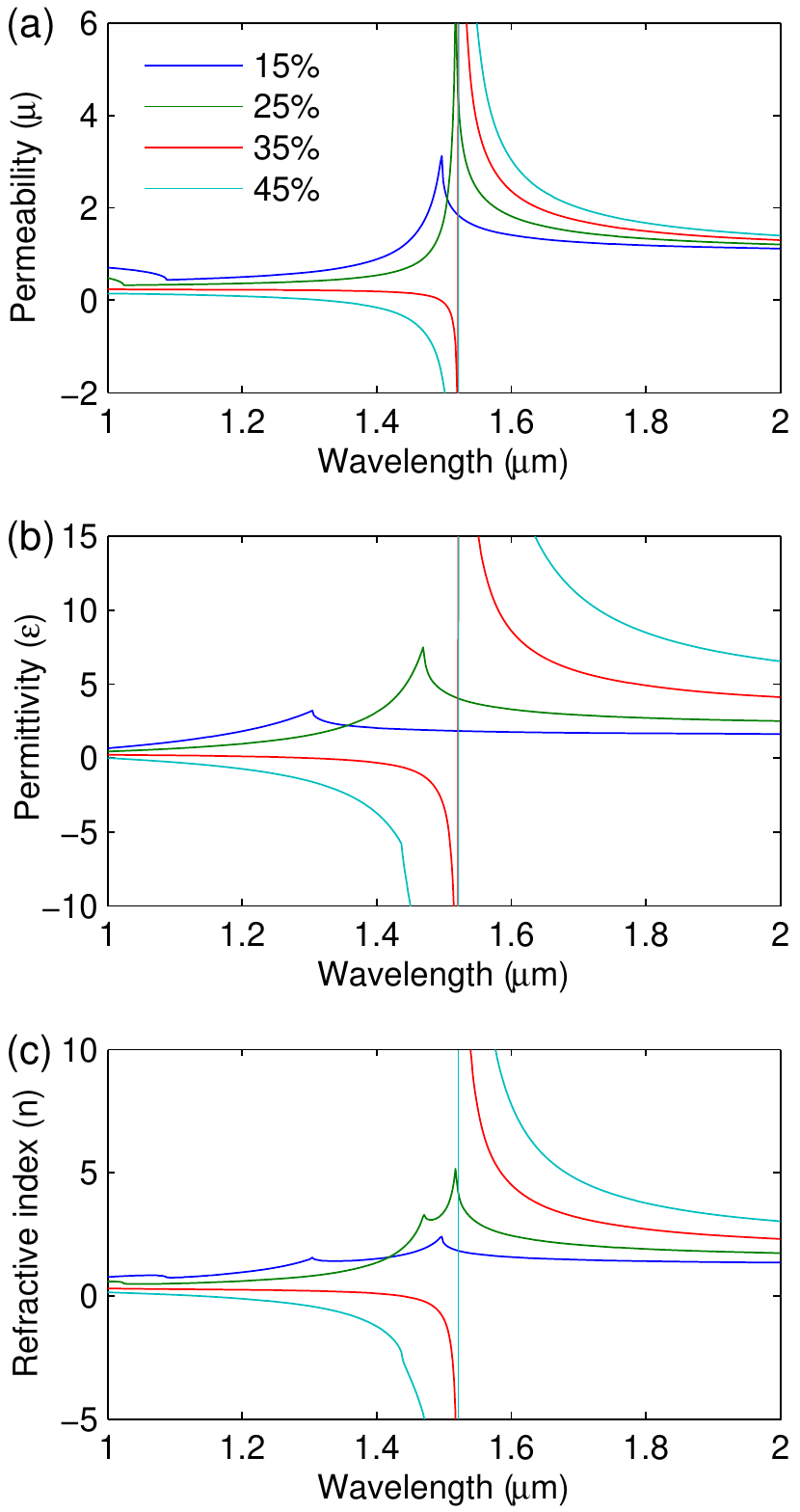}
\caption{\label{fig:epsart} Wavelength dependence of the real part of the effective permeability (a), permittivity (b), and refractive index (c) for 380 nm diameter Si spheres randomly dispersed in free space for different volume loading fractions.}
\end{figure}

\begin{figure}
\includegraphics[width=47mm]{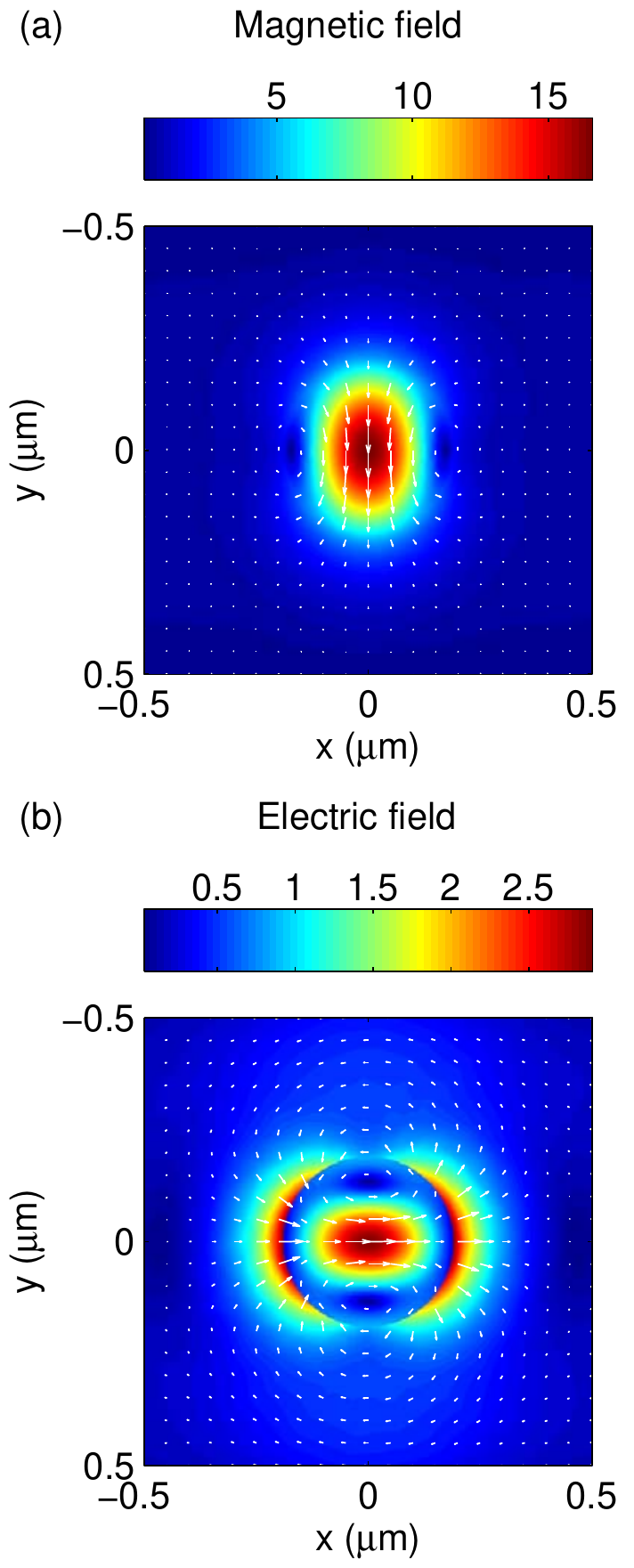}
\caption{\label{fig:epsart} Magnetic (a) and electric (b) fields, normalized to the incident fields, at their respective Mie resonances for 380 nm diameter Si spheres in free space. The incident electric field is along x and propagation is into the page.}
\end{figure}

As an example of how percolation and Mie resonance induce a negative index, consider a metamaterial operating in the near infrared band consisting of 380 nm silicon (Si) spheres dispersed in free space. Silicon is chosen for its large permittivity ($\epsilon_p\approx12$), which is a requirement to obtain strong Mie resonances. The calculated wavelength dependence of the effective permeability, permittivity, and refractive index for different volume loading fractions, calculated using Eqs. (10) and (11), are shown in Fig. 1. For a relatively low loading of 15\%, the magnetic-dipole Mie resonance gives rise to a resonance in the permeability at 1.5 $\mu$m, while the electric-dipole Mie resonance leads to a resonance in the effective permittivity at 1.3 $\mu$m. For this loading, the permeability and permittivity are both positive throughout the band. As the loading increases to 25\%, the amplitude of both resonances increases, while the electric resonance in the permittivity shows a significant spectral redshift. Again, both parameters are positive for this loading. As the loading increases to 35\%, which exceeds the percolation threshold of 33\%, the redshift of the electric resonance causes it to overlap with the magnetic resonance. Since the permeability and permittivity are both negative near the resonance, the effective refractive index is also negative in the overlap region around 1.5 $\mu$m. When the loading is increased further beyond percolation to 45\%, the resonances remain overlapped and the index remains negative. Away from the resonances, the loading dependence of $\epsilon$ and $\mu$ is similar to previous models \cite{Lewin1947,Wu2006,Slovick2014}. Finally, to confirm that the effective parameters in Fig. 1 are analytic functions of the wavelength, the Kramers-Kronig relation was applied to their corresponding imaginary parts and found to be equivalent to the values shown.

The reason why only the electric resonance undergoes a significant redshift can be understood from the electric and magnetic field distributions around the spheres at their respective resonances, shown in Fig. 2. While the magnetic field (\textbf{H}) is well confined at the magnetic resonance (1.4 $\mu$m), the electric field (\textbf{E}) at resonance (1.04 $\mu$m) shows considerable enhancement outside the sphere. The different degrees of confinement of the electric and magnetic fields can be traced to their different boundary conditions. Since the normal components of \textbf{B} (=$\mu$\textbf{H}) and \textbf{D} (=$\epsilon$\textbf{E}) must be continuous across the sphere surface, the normal component of \textbf{H} is also continuous because $\mu=1$ throughout, whereas the normal component of \textbf{E} is discontinuous owing to the large dielectric mismatch between Si and free space. This leads to a larger normal component of $\mathbf{E}$ just outside the sphere surface, and hence poor confinement. Alternatively, the normal component of $\mathbf{H}$ is continuous, leading to better field confinement. The poor confinement of the electric field leads to interparticle coupling at high loading, lowering the energy of the electric dipole-dipole interaction, and redshifting the resonance.

A drawback of disordered metamaterials is that their lack of periodicity leads to scattering loss, which limits the thickness of useful material to approximately one mean free path. A lower limit on the mean free path can be obtained using the single scattering model. In contrast to the effective medium, the single scattering model does not account for scattering cancellation by the matrix, and thus predicts the maximum extinction. Figure 3 shows the calculated single-scattering mean free path, normalized to the particle diameter, for the material in Fig. 1. As expected, the mean free path decreases at shorter wavelengths, reaching a local minimum near the magnetic Mie resonance at 1.4 $\mu$m. For 35\% loading, the mean free path in the negative index region around 1.5 $\mu$m is approximately equal to the particle diameter. Thus, to obtain acceptable transmission in the disordered material with negative index, the thickness should be limited to approximately one unit cell.

\begin{figure}
\includegraphics[width=65mm]{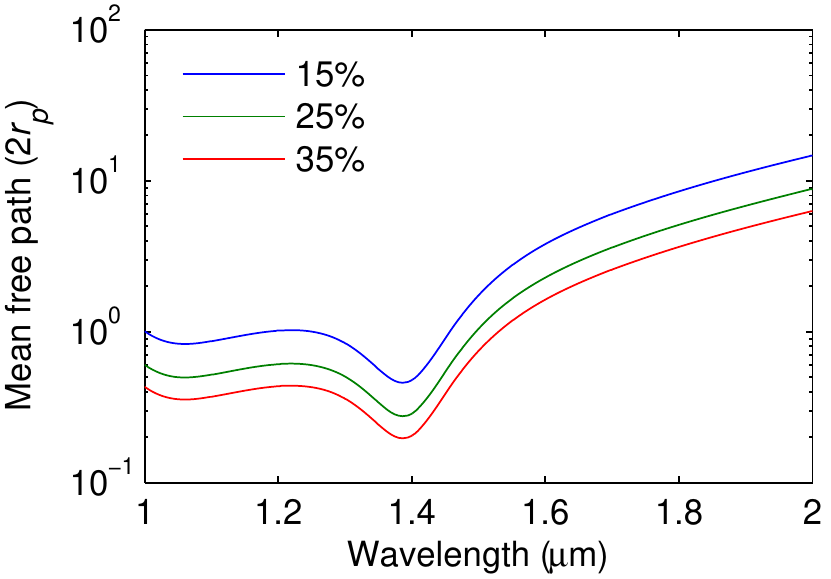}
\caption{\label{fig:epsart} Wavelength dependence of the mean free path for 380 nm diameter Si spheres randomly dispersed in free space for different volume loading fractions.}
\end{figure}

In summary, the classic Bruggeman percolation model for disordered composites was generalized to support particle sizes comparable to the wavelength. Similar to effective medium models for ordered metamaterials, this model predicts that electric and magnetic Mie resonances in the spheres give rise to resonances in the effective permittivity and permeability of the composite. The model also predicts a redshift of the electric resonance with increasing particle loading due to particle interactions arising from poor confinement of the electric field. Most importantly, when the particle loading exceeds the percolation threshold of 33\%, the electric resonance coincides with the magnetic resonance, resulting in a negative refractive index. The possibility of obtaining a negative index in disordered materials opens the door to more scalable designs suitable for large-area applications.

\bibliography{bib}

\end{document}